\documentclass[11pt]{article}
\usepackage[margin=1.0in]{geometry}
\usepackage{graphicx}
\usepackage{epsfig}
\usepackage{amssymb,amsfonts,amsmath}
\usepackage{mathrsfs}
\usepackage{epstopdf}
\usepackage{hyperref,color}
\usepackage{natbib}
\usepackage{setspace}

\begin{document}

\title{The fate of presentism in modern physics}
\author{Christian W\"uthrich\thanks{I am indebted to Ad\'an Sus and Giuliano Torrengo for helpful comments on an earlier draft. Work on this project has been supported in part by the American Council of Learned Societies through a Collaborative Research Fellowship, the University of California through a UC President's Fellowship in the Humanities, and the University of California, San Diego through an Arts and Humanities Initiative Award.}}
\date{Forthcoming in Roberto Ciunti, Kristie Miller, and Giuliano Torrengo (eds.), {\em New Papers on the Present---Focus on Presentism}, Philosophia Verlag, Munich.}
\maketitle

\begin{abstract}\noindent
There has been a recent spate of essays defending presentism, the view in the metaphysics of time according to which all and only present events or entities exist. What is particularly striking about this resurgence is that it takes place on the background of the significant pressure exerted on the position by the relativity of simultaneity asserted in special relativity, and yet in several cases invokes modern physics for support. I classify the presentist arguments into a two by two matrix depending on whether they take a compatibilist or incompatibilist stance with respect to both special relativity in particular and modern physics in general. I then review and evaluate what I take to be some of the most forceful and intriguing presentist arguments turning on modern physics. Although nothing of what I will say eventuates its categorical demise, I hope to show that whatever presentism remains compatible with empirical facts and our best physics is metaphysically repugnant.
\end{abstract}

\section{Introduction: ersatzist presentism}\label{sec:intro}

Defining `presentism' in a way that saves it from being trivially false yet metaphysically substantively distinct from eternalism is no mean feat, as the first part of this collection testifies. In \cite{wut11}, I have offered an attempt to achieve just this, arguing that this is best done in the context of modern spacetime theories. Here, I shall refrain from going through all the motions again and simply state the characterization of an ersatzist version of presentism as it has emerged from considerations there. Any acceptable formulation of presentism should remain neutral among competing spacetime theories in order to enable the present project of assessing the compatibility of presentism with various theories of modern physics, including both spacetime theories and theories of physical processes situated in a spatiotemporal setting. 

The main issue in the triviality debate as I see it concerns the representation of events without an accompanying ontological commitment. If the presentist can find a way to represent non-present events without {\em eo ipso} committing herself to their existence, then expressing her metaphysical disagreement with the eternalist seems rather straightforward. This naturally leads to an ersatzist position which introduces non-present events merely for representational purposes without imbuing them with physical existence.

The vantage point of modern spacetime theories is the presupposition of a four-dimensional manifold $\mathcal{M}$ with certain topological and differential structure. Furthermore, the manifold $\mathcal{M}$ is equipped with a metric field $g_{ab}$ which encodes all the information concerning the spatiotemporal relations among all the points of $\mathcal{M}$. Eternalism and presentism are then taken to disagree as to over which points of $\mathcal{M}$ they quantify when quantifying over all spatiotemporal events where physically existing entities can be located. In this context, {\em eternalism} is understood as the position claiming physical existence for all events in $\mathcal{M}$. In contrast, {\em presentism} partitions $\mathcal{M}$ into past, present, and future events. This partition results, e.g., from assuming an equivalence relation $S$ (`simultaneity') to be defined on $\mathcal{M}$ such that the equivalence classes contain co-temporal events. Time, on this view, is the one-dimensional linearly ordered quotient set of these equivalence classes. One such equivalence class is privileged in that it contains the `present' events, the set of equivalence classes to its past according to this ordering contains the `past' events and the set of equivalence classes to its future the `future' events. Obviously, the sets of all past and future events thus have further structure indicating just how much to the past or future a particular event is located. Thus, the sum total of physical existence according to the presentist is a proper subset of that according to the eternalist. 

An obvious worry arising from this manner of characterizing the position is that presentism does not just amount to the assertion that only present events or entities exist, but also that the present undergoes a dynamical `updating', or exhibits a quality as of a fleeting whoosh, and that this additional dynamical aspect is what threatens the substance of the debate between the presentist and her eternalist opponent. In order to capture this dynamical quality, the thought goes, the presentist must quantify not just over the events contained in one equivalence class corresponding to the present present, but also over all events in all the other equivalence classes containing the past and future presents. Once this point is granted, it seems as though presentism deflates into admitting all the events of $\mathcal{M}$ as existing. But this clearly misses the presentist's point: the presentist's sum total of existence remains a proper subset of the eternalist's, fleeting whoosh or not. I am not pretending as if to characterize in satisfactory detail what exactly constitutes this dynamical quality is without difficulty. But for present purposes, presentism should be understood as a merely {\em ontological} hypothesis making an assertion as to what exists, and not an {\em ideological} statement about the qualities---dynamical or otherwise---of that which exists. Perhaps this is a mistake. But if it is, at least not without precedent.

The remainder of this essay shall assume, however fallibly, that presentism is a metaphysically substantive thesis markedly different from eternalism. It contends that physical existence is restricted to a spatially extended manifold of events simultaneous with the {\em here-now}. This view comes under severe pressure from modern physics, most notably from special relativity (SR), as shall be explicated in Section \ref{sec:chall}. The source of the tension is found in the fact that in SR, and hence in modern physics, space and time are intertwined in a way such that whether two given spacetime events exemplify the relation of simultaneity is no longer an absolute and global matter. But if simultaneity cannot serve as on absolute and global basis for determining whether or not a spatially distant event is present (in the temporal sense), then we seem to lack an objective basis on which matters of physical existence could turn for a presentist metaphysic. 

Naturally, presentists have responded to the challenge. The problem, of course, should not be misconstrued as dealing with an in principle insurmountable inconsistency between presentism and physics; rather, the challenge amounts to grounding the necessary distinctions (past, present, and future) in a way that is responsive to modern physics while remaining faithful to presentist intuitions. The presentist responses to this challenge, both actually stated and hitherto unarticulated, shall be chronicled in the remainder of this essay, together with an assessment of the prospects of success and the price tag for each response. In an attempt to bring order into the variegated multitude of presentist strategies to counter the challenge from modern physics, a systematization is offered in Section \ref{sec:taxon}. The basic distinction of presentist responses is into compatibilist and incompatibilist strategies, with the former arguing that presentism is compatible with the truth of SR despite initial appearances and the latter accepting their incompatibility while rejecting that this entails the denial of presentism. It turns out to be useful to introduce a distinction orthogonal to the one between compatibilism and incompatibilism: presentism can not only be compatible or incompatible with respect to SR, but to fundamental physics---contemporary or prospective. This distinction derives its utility from the fact that there are a number of presentist retorts readily admitting that their view is inconsistent with SR but insisting on its compatibility with fundamental physics. 

These two distinctions span a matrix of four types of strategies. The two boxes of strategies accepting an incompatibility with contemporary, and possibly future, fundamental physics will be examined in Section \ref{sec:taxon}. Responses purporting a compatibility with either contemporary or at least future fundamental physics will be dealt with in Sections \ref{sec:com}, if they also allege a compatibility with SR, and \ref{sec:incom}, if they accept that presentism is inconsistent with SR. Section \ref{sec:stock} will take stock and dare a rather negative comprehensive appraisal of the prospects of presentism to survive the pressure from modern physics in any form that permits retaining its appeal.

\section{The challenge issued by special relativity}\label{sec:chall}

If by {\em presentism} we thus mean the thesis according to which there exists an absolute spatially extended present and all there is is spatiotemporally located in this present, then a strong argument can be offered to the effect that such a position is precluded if SR is at least approximately true. The goal of this section is to carefully develop this argument.

Starting out from two major premises,\footnote{And some minor ones, such as the homogeneity and isotropy of space and time. The two major axioms of SR are the {\em Light Postulate}, according to which ``light propagates through empty space with a definite velocity which is independent of the state of motion of the emitting body'' \citep[891]{ein05}, and the {\em Relativity Principle}, according to which ``the same laws of electrodynamics
and optics will be valid for all frames of reference for which the equations of mechanics hold good [i.e., for inertial frames]'' (ibid.).} SR asserts a certain structure of space and time. In 1908, Hermann Minkowski showed that this inferred structure is best captured by postulating a four-dimensional manifold of `events', i.e.\ of dimensionless points, which is differentiable and endowed with the additional structure of a time orientation and a metric field encoding the absolute spatiotemporal---but not the spatial or temporal---separation between events. In fact---and this cuts to the core of the difficulty for the presentist---, there simply {\em is} no absolute spatial or temporal measure of separation in SR. In Minkowski's famous words: ``Henceforth space by itself, and time by itself, are doomed to fade away into mere shadows, and only a kind of union of the two will preserve an independent reality.'' \citep[75; translation of \citet{min08}]{min52} What is absolute, i.e.\ independent of a frame of reference, is only the {\em union} of space and time; in contrast, the totality of space at a particular time is only even defined relative to a frame of reference. This means that there simply is no absolute and objective truth concerning which spatially distant events are simultaneous with the event representing the here and now.

\begin{figure}
\centering
\epsfig{figure=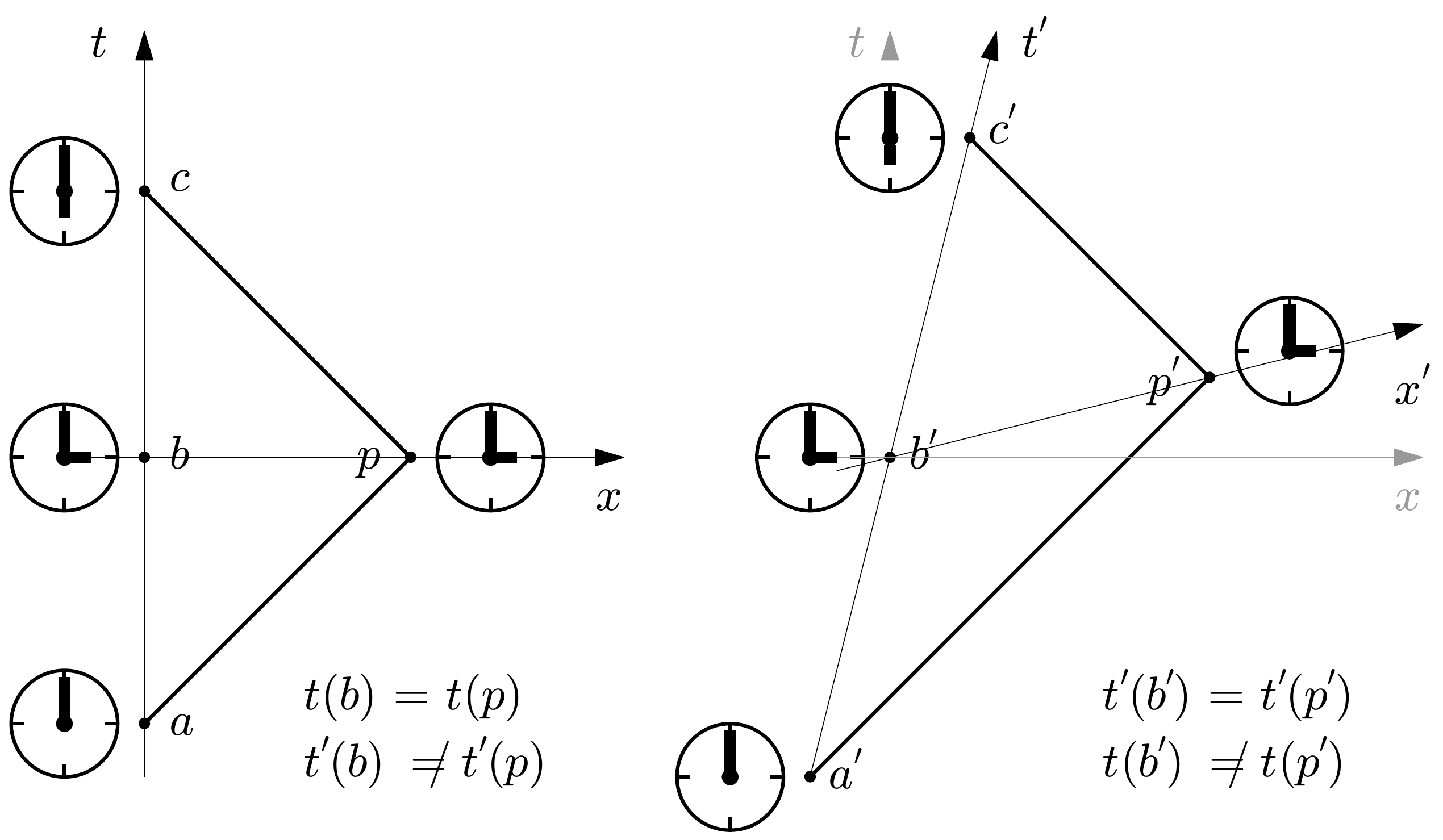,width=0.8\linewidth}
\caption{\label{fig:clocksync} Einstein-Poincar\'e convention for synchronizing distant clocks in two reference frames related by a Lorentz transformation}
\end{figure}

Let us state this with slightly more rigour. From the two premises mentioned, it follows that the uniquely correct way to transform the time and space coordinates of events---and hence the assignment of temporal and spatial location---of two different inertial frames in relative motion is by employing so-called `Lorentz transformations'. Somewhat imprecisely, Lorentz transformations are a kind of hyperbolic rotation in a mathematical space including a time direction.\footnote{For an intuitive derivation and illustration, \citet{janxx} is highly recommended. \Citet{giu10} is a good source for the more technical aspects of Minkowski spacetime.} Apart from all the other fun consequences of Lorentz-transforming physical systems such as time dilation, length contraction, and the infamous twin paradox, Lorentz transformations also have profound consequences for the temporal (and spatial) ordering of events. To see how Lorentz transformations affect this ordering, consider the so-called {\em Einstein-Poincar\'e convention} for synchronizing spatially distant clocks by means of light rays, illustrated in Figure \ref{fig:clocksync}.
According to this convention, a spatially distant point $p$ is simultaneous with an event $b$ in the rest frame of a clock just in case a light ray is reflected back to the clock at $p$ such that the same duration lies between the events $a$ of the emission of the light ray and $b$ as does between the event $c$ of the reception of the light ray and $b$. In other words,
\begin{equation*}
t(b) = t(p) \Leftrightarrow t(b) - t(a) = t(c) - t(b),
\end{equation*}
where $t(e)$ gives the time coordinate of any event $e\in \mathcal{M}$ in the unprimed reference frame. If we Lorentz transform into another reference frame---with transformed coordinates or manifold points denoted by primes---and apply the same convention to determine the set of spatially distant events that are simultaneous with $b\equiv b^\prime$, it becomes evident that the set of spatially distant simultaneous events is different in different reference frames. For starters, the point $p$ on the unprimed $x$-axis is not simultaneous with $b\equiv b^\prime$ according to the primed frame:
\begin{equation*}
t^\prime(b) 
\neq t^\prime (p).
\end{equation*}
Instead, the primed frame evaluates the event $b\equiv b^\prime$ as simultaneous with an event $p^\prime$, which is later than $p$ (in both frames) and hence not simultaneous with $b$ according to the unprimed frame:
\begin{eqnarray*}
t^\prime(b^\prime) &=& t^\prime (p^\prime),\\
t(b^\prime) &\neq& t(p^\prime).
\end{eqnarray*}
In short, for an inertial observer at rest with respect to the {\em unprimed} frame, all the spacetime events on the $x$-axis are simultaneous with $b$, whereas for an inertial observer at rest with respect to the {\em primed} frame, those spacetime events on the $x^\prime$-axis are simultaneous with $b^\prime \equiv b$. Hence, the simultaneity of distant events is no longer absolute but only relative to inertial frames once one accepts the Lorentz symmetry demanded by SR.\footnote{One might be tempted to think that the problem only arises because simultaneity is {\em conventional}, as was implied above when I started out from the Einstein-Poincar\'e {\em convention} of simultaneity. But this would be missing the point: while different conventions are surely possible, the Einstein-Poincar\'e convention is uniquely suited for presentist purposes as it maintains the symmetry of simultaneity (within a fixed frame) by choosing the midpoint between $a$ and $c$, while other conventions would not help against the {\em relativity} of simultaneity and make things worse by frivolously giving up the symmetry of the relation. In fact, non-standard conventions of simultaneity could threaten presentism already in pre-relativistic physics. But this threat could easily be averted by choosing a sensible (standard) convention of what it is to be simultaneous.}

The presentist asks us to be realist about all events and objects in the {\em present}, but no others. Lest the position collapses to a solipsistic denial of the reality of anything at a spatial distance, she thus needs to procure an account of what we are to include among the things present---and to exclude as not being part of the present. In other words, presentism must involve an at least implicit commitment to a way to determine the simultaneity, and hence co-presence, of spatially distant events with the vantage point of the here and now. Prior to the advent of relativistic physics, such a commitment was both unambiguous and unproblematic insofar as pre-relativistic physics readily offered a robust notion of absolute simultaneity. But relativity appears to pull the rug from underneath any metaphysical view which relies on an objective, i.e.\ absolute, determination of what is past, present, and future. 

An argument against metaphysical views of time that postulate or entail that the future is genuinely open in the sense that it is not (yet) real, or does not (yet) exist, as of the present moment has been advanced by Wim \citet{rie66} and Hilary \citet{put67}.\footnote{The sense in which the future is supposed to be genuinely open is important to note, as eternalism is arguably consistent with at least some forms of indeterminism. For a reading of Rietdijk's and Putnam's argument as an argument to the conclusion that SR is deterministic, see \citet[\S 4.3]{rak97}.} This argument starts out from the assumption that the task at hand is to determine the set of spatially distant spacetime events which are simultaneous, and hence {\em co-present}, with the {\em here-now}, the vantage point from which the present is thus to be constructed. As a next step, invoke the equivalence relation $S$ which we found in \S\ref{sec:intro} the presentist to rely on. Physically, this binary relation is interpreted to signify the simultaneity between two spacetime events. Mathematically, it enables the partitioning of the spacetime events into equivalence classes of events, ordered by a time parameter. Metaphysically, it creates the sets of co-existing events. On a presentist metaphysics, to repeat, one of these equivalence classes is privileged in that its elements alone exist concretely. In these terms, the task can be characterized as that of being handed an event representing the {\em here-now} and a binary relation $S$ which we are to use to determine which other events exist. 

\begin{figure}
\centering
\epsfig{figure=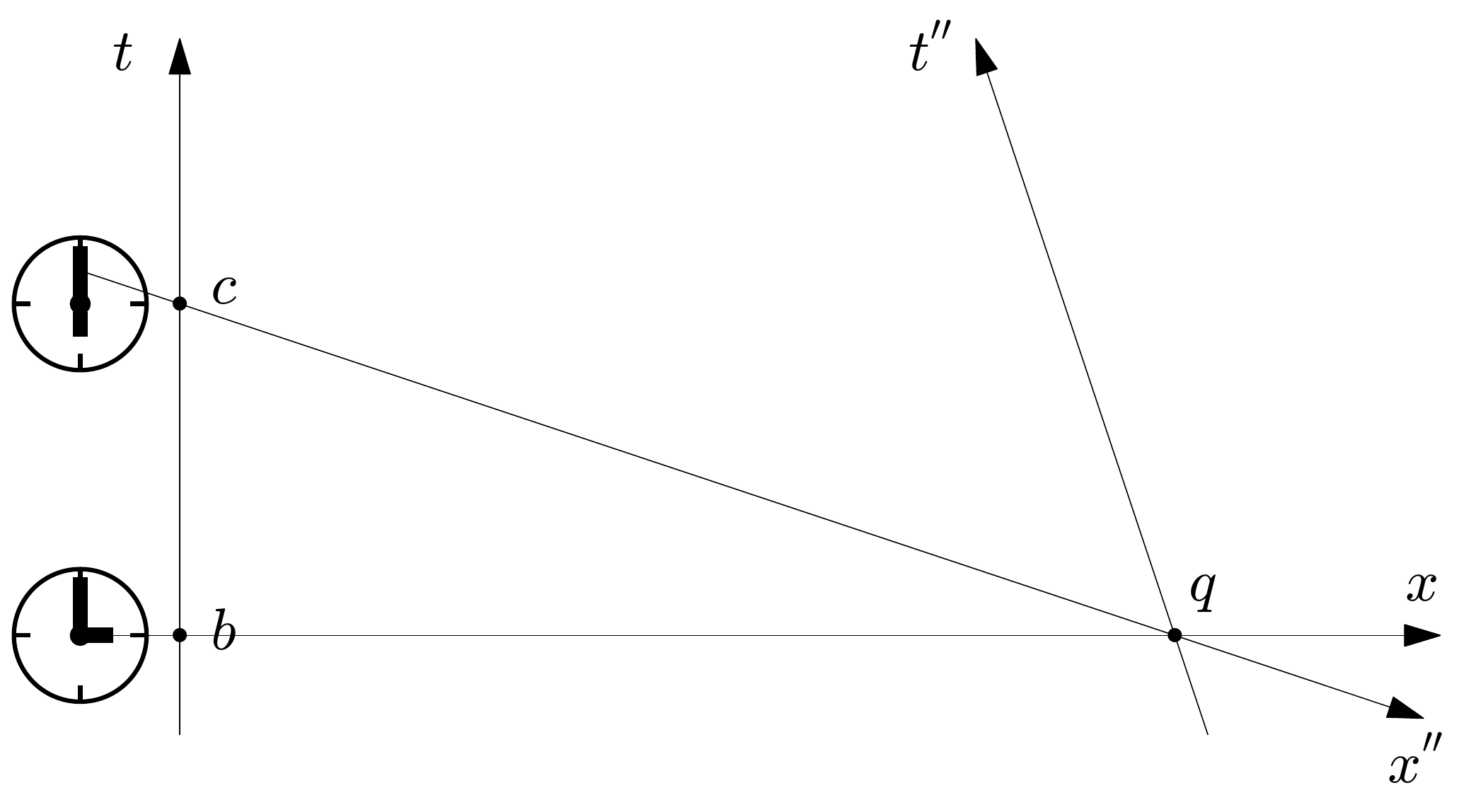,width=0.7\linewidth}
\caption{\label{fig:transitivity} What the transitivity of simultaneity can do: $Sbq$ in the unprimed frame, $Sqc$ in the doubly primed frame}
\end{figure}

As a consequence of the relativity of simultaneity found in SR, if event $b$ denotes the {\em here-now} as in Figure \ref{fig:clocksync}, there is no frame-independent way to determine the set of events that stand in relation $S$ to $b$. As far as the unprimed frame of reference is concerned, we have $Sbp$. In the primed frame, however, we find that $\neg Sbp$ but $Sbp^\prime$. Since $S$ is an equivalence relation, and hence transitive, whichever events stand in $S$ to events which stand in $S$ to $b$ should also stand in $S$ to $b$. If the qualification that whether two events stand in $S$ or not can only be determined with respect to a frame of reference is omitted, then the transitivity of $S$ seems to entail that, since there exists an event $q$ such that $Sbq$ (in some frame) and $Sqc$ (in some (other) frame), it is the case that $Sbc$. See Figure \ref{fig:transitivity} for an example of such an event $q$, with $b$ and $c$ related as in the left-hand side of Figure \ref{fig:clocksync}. But this is absurd: $S$ is supposed to be a relation of simultaneity, yet $c$ is clearly to the future of $b$! That $c$ is to the future of $b$, importantly, is frame-independent and hence agreed on by all inertial observers. But if it is absolutely and objectively the case that $c$ is to the future of $b$, they cannot stand in any relation that can sensibly be interpreted as a relation of simultaneity. 

Note that the absurdity is rampant: for any pair of events $a$ and $b$ in the manifold $\mathcal{M}$ of Minkowski spacetime, there exists an event $c\in\mathcal{M}$ such that $Sac$ and $Sbc$ and hence $Sab$. Hence, $S$ is the universal binary relation on the set of events $\mathcal{M}$. But surely a presentist would not want to be bound by an ontological commitment to all events in spacetime. From this consequence, both Rietdijk and Putnam have concluded that any metaphysical position marking ontological distinctions along a relation of simultaneity is thus reduced to absurdity.

Of course, one might interject that to let the transitivity of $S$ act across different reference frames is illicit; the central lesson of the relativity of simultaneity in SR is that such transitivity only obtains within the same reference frame. In fact, while simultaneity remains an equivalence relation, it might be argued, once we accept the Lorentz symmetry of SR it does so only within each frame. If this point is heeded, it might then be concluded, the argument above no longer goes through. True, but to concede that the transitivity ought to be restricted to within the same reference frames and hence that simultaneity is equally restricted to reference frames surely seems to concede too much as far as the presentist is concerned. After all, presentism relies, it seems, on an absolute notion of simultaneity in order to make absolute ontological claims.

Without going into the details of presentist responses to this challenge just yet, let us also note that it won't suffice for the presentist to merely reject the way of constructing a spatially extended present as offered in this section. The presentist might be tempted to argue that the argument presented here does not even get to the starting block as the idea of starting out from some privileged or arbitrarily chosen spacetime event (the `here-now') and then trying to identify those events simultaneous to it. Instead, she might be tempted to think, the present and what is contained in it is primitively given, it is there `at once', prior to us doing any physics. But suppose that's the case. It would then still be true that, if you hand me just one event as being an element of the present (or, {\em eo ipso} according to the presentist, of physical reality), the set of all primitively given elements of the present would form a three-dimensional submanifold of Minkowski spacetime containing the one starting point. I don't see how this move does not amount to privileging one particular way of carving up spacetime into equivalence classes of simultaneous events and, furthermore, of privileging one particular such equivalence class as the `present', be that primitive or not. 

The presentist might retort that this way of conceiving of the problem does not get started if we don't help ourselves to this one event from which we subsequently try to construct the rest of the present. But surely, she might continue, the positing of such a vantage point is wrongful, at least without some further motivation. True, if this opening move is barred, the challenge can't be constructed as above. But I fail to see what the presentist could win from disallowing it. She would claim, in essence, that all and only present events and objects are part of physical reality {\em and} that there is in principle no way of determining even one element of this physical reality. Lest we permit ourselves to lapse into obscurantism, the presentist ought to accept the challenge as it stands---particularly given the plethora of more interesting responses available to her.

\section{A taxonomy of presentist responses to the challenge}{\label{sec:taxon}

Let us then consider and classify actual, and possible but unstated, presentist responses to the argument as outlined in Section \ref{sec:chall}. The basic distinction I wish to use in systematizing presentist reactions is that between compatibilism and incompatibilism between presentism and SR. Versions of {\em compatibilism} with SR assert that, despite appearances, SR and presentism are perfectly compatible in that they can jointly and consistently be maintained. Juxtaposed, we find varieties of {\em incompatibilism} with SR, which accept the argument as given in \S\ref{sec:chall}, but reject that it entails the denial of presentism. Clearly, then, an incompatibilist of this kind is thus obliged to reject SR. For some, such a move is justified on the background of their rather sweeping rejection of physics as a science whose task it is to unveil facts about our physical world that a philosopher ought to take into account when constructing metaphysical theories. But there are others in this camp who, while rejecting SR, are adamant about dismissing such an attitude of wholesale rejection of physics as being irrelevant to the task at hand. For them, a metaphysics blatantly contradicting our best physical theories is indefensible. An incompatibilist of this sort, then, has to deny that SR, at least as standardly understood, is among our best physical theories.\footnote{Note that, as will subsequently become clear, what exactly SR is taken to assert or entail will be of paramount importance when judging the (in)compatibility of presentism with it.} 

In order to make room for this additional distinction, it seems sensible to distinguish between more encompassing forms of both compatibilism and incompatibilism, not just with SR, but with modern physics {\em in toto}, i.e., with contemporary, or in fact prospective, fundamental physics. Since it is at least logically possible to sever the two distinctions and be, e.g., an incompatibilist with respect to SR but not modern physics in total, they are strictly speaking orthogonal and give rise to a two-by-two matrix of four distinct types of presentist strategies in the face of the challenge mounted in \S\ref{sec:chall}, as follows:

\begin{table}[h]
\centering
\begin{tabular}{l||l|l}
 & compatibilism with SR & incompatibilism with SR\\ \hline\hline
compatiblism with & presentism compatible with & presentism incompatible with\\ 
modern physics & both SR and physics & SR, but compatible with physics \\ \hline
incompatibilism with & presentism compatible with SR, & presentism incompatible with \\ 
modern physics & but incompatible with physics & both SR and physics
\end{tabular}\caption{\label{tab:matrix} The matrix of distinct types of presentist strategies}
\end{table}

While all four options are logically possible, it is evident that they are not all equally attractive. The lower left box in Table \ref{tab:matrix}, for instance, has not been defended in print, to the best of my knowledge. This is hardly surprising, for why should a presentist go at any length establishing compatibility with SR, only to then concede that it remains incompatible with fundamental physics. If a defender of presentism estimates her theory to be in conflict with fundamental physics, why spend any effort to defend its consonance with SR? Such a strategy would therefore only appear to be rational, it seems, if one believed that the irreconcilability of presentism with other theories in physics can be dispelled in ways that one with SR could not. But there is no reason to believe that {\em that} is the case: as will become clearer below, the reason why theories in fundamental physics clash with presentism, if any, is that we ask them to respect the Lorentz symmetry demanded by SR. In a sense, then, the conflict arises because of, and to the extent to which, fundamental physics is required to be special-relativistic. It seems odd, then, to argue for a compatibility of presentism and SR, while maintaining a discordance between presentism and other theories in fundamental physics. Consequently, I will not consider potential presentist responses that would fall in the lower left quadrant of Table \ref{tab:matrix} any further.

What about the box on the lower right-hand side? The acceptance of a conflict between presentism and not only SR, but all of current, as well as prospective, fundamental physics paired with an insistence on presentism amounts to a rather comprehensive rejection of physics. It thus fundamentally contravenes naturalism, a venerable tradition going back at least to Aristotle. According to naturalism, philosophical---and metaphysical---inquiry is continuous with scientific inquiry. To be sure, naturalism is not a logical truth---it is a substantive philosophical thesis. But it is one whose defence has to wait for another day; for present purposes, I simply assume a minimal naturalism which demands that no philosophical thesis be in manifest contradiction to facts established by our best science. Restricting this weak thesis to metaphysics, it can be translated as necessitating that the physically possible worlds are a subset of the metaphysically possible ones, for if the metaphysical theories were in contradiction to the physical ones, then there would have to be some physically possible worlds (and perhaps all) which are metaphysically impossible, as for the metaphysical theory to be incompatible with physics, it would have to rule out some physically possible worlds as impossible.\footnote{Of course, this also presupposes that the ``facts established by our best science'' get translated as those facts compatible with the laws of our best physical theories.} In other words, metaphysics would a priori deem impossible what physics affirms is possible. Assuming that all physically possible worlds are also logically possible, I see little justification for disavowing this weak form of naturalism. 

In what follows, I shall hence assume that the most attractive presentist strategies are to be found in the camp espousing compatibilism with fundamental physics. This leaves us with the top two boxes of Table \ref{tab:matrix}, and thus with either compatibilism or incompatibilism with respect to SR.

\section{Compatibilism with special relativity}\label{sec:com}

There are various ways in which one could work out a compatibilist response (regarding both SR and physics in general). An obvious way to do so would be to accept a modification of the presentist position such that the reformulated thesis is compatible with Lorentz symmetry. Although this would not by itself guarantee that the reformulated position is compatible with any future fundamental physical theory, it would remove any immediate reason for believing that it couldn't be. Of course, given that we do not currently have at our disposal the final and true fundamental theory, it would be illusory to seek such a guarantee. Thus, a compatibilist must content herself with making an informed bet on which parts of our current physics are likely to be retained, in a sufficiently similar form, in the final theory. Accordingly, the modifications required for compatibilism can only conform to what our currently best judgments concerning this are. 

Apart from modifying the presentist thesis, there are, broadly speaking, at least two further ways for the full compatibilist to work out an answer. As a second option, one can argue that SR, and any other relevant physical theory, are not about time, or at least not about the same sort of time as the presentist is concerned with. Since their objects are thus distinct, there could not possibly be an inconsistency between presentism and physical theories. Hence, they are perfectly compatible; and since this reasoning applies to any future physical theory, this argument concludes, we can remain happy compatibilists until the end of time. 

The third---and surprisingly popular---option denies that SR, properly interpreted, involves or entails an assertion to the effect that there cannot be any absolute, i.e.\ observer-independent, simultaneity relation $S$. In fact, proponents of this strategy insist, what SR does prohibit is only that any such absolute simultaneity could not be detected in principle and would hence remain empirically completely inaccessible. Thus, SR does not preclude the existence of an absolute, non-empirical $S$. Since such an $S$ does exist, though undetectably so, there is no problem in identifying the spatially distant events which are co-present with the {\em here-now}. To be sure, this identification cannot be executed in practice, as $S$ must remain behind a principled veil of ignorance, but the possibility that it exists assures the presentist that there can be a privileged simultaneity relation and thus an objectively distinguished present. So if SR is interpreted as to only imply that there cannot exist an absolute $S$ which can also be detected, but not to entail that there cannot be an absolute non-empirical $S$, then presentism remains compatible with SR and arguably with all of physics.

In sum, then, it appears as if the compatibilist can select among three different routes: either insist that SR and presentism talk about different things and hence circumvent the issue of compatibility, or modify presentism such as to eliminate any tension with SR, or re-interpret---and arguably modify---SR such that it no longer entails that there cannot be absolute non-empirical simultaneity. Let me discuss these options in some more detail.

The first road claims that SR, unlike presentism, is not really a theory about `time', in spite of any appearances to the contrary. Perhaps the most prominent proponent of this view was Arthur Prior. We find the clearest expression in one of his posthumously published essays:

\begin{quote}
[W]e may say that the theory of relativity isn't about {\em real} space and time, in which the earlier-later relation is defined in terms of pastness, presentness, and futurity; the `time' which enters the so-called space-time of relativity theory isn't this, but is just part of an artificial framework which the scientists have constructed to link together observed facts in the simplest way possible, and from which those things which are systematically concealed from us are quite reasonably left out. \citeyearpar[50f; emphasis in original]{pri96}
\end{quote}

Prior claims, in effect, that the `time' in SR is of merely instrumental value, used in physics as an ordering parameter of in principle observable events. Real time (and space) which are defined relationally in terms of pastness, presentness, and futurity, he implies, is systematically concealed from us, as of course it has to on pains of violating the Lorentz symmetry demanded by SR. His implication that there is, ontologically speaking, an absolute and objective fact of the matter where events stand in terms of their pastness, presentness, and futurity, even though this fact must remain forever invisible to us, comes awfully close to the third way of giving a compatibilist response as I have sketched it above. In fact, more than a score years earlier, though still posthumously, Prior wrote that

\begin{quote}
[o]ne possible reaction to this situation, which to my mind is perfectly respectable though it isn't very fashionable, is to insist that all that physics has shown to be true or likely is that in some cases we can never {\em know}, we can never {\em physically find out}, whether something is actually happening, or merely has happened or will happen. \citeyearpar[323; emphases in original]{pri72}
\end{quote}

It is obvious why this view is not as fashionable as perhaps Prior would have hoped (although it's still surprisingly popular): it constitutively asserts what cannot be known. Even though Prior seems insufficiently impressed by this principled ignorance---he believes it to only apply to ``some cases''---, it is important to emphasize just how generic it is: although we can determine events in the past lightcone of the {\em here-now} to be past as causal signals emanating from them can in principle reach the {\em here-now}, no spatially distant event can ever be known, or ``physically found out'', to be present. The only event of which we can ascertain its presentness, and hence, according to the presentist, its very existence, is the {\em here-now}. Hence, the principled epistemic strictures imposed by SR are much more constraining than Prior seems to realize. 

It should also be stressed that Prior seems to accept the challenge as it has been set up in \S\ref{sec:chall}, as whether a spacelike related event is co-present with the {\em here-now}, ``[o]n the view of presentness which [he has] been suggesting, this is {\em always} a sensible question.'' \citep[322; emphasis in original]{pri72} The task, according to him, is exactly to identify a relation not just of simultaneity with respect to a frame of reference, but of simultaneity {\em tout court}. Thus, Prior accepts the challenge as it stands and appears to vacillate in his response between saying that SR and presentism refer to different things when they state `time' and thus cannot be incompatible, and saying that SR leaves open the possibility of an absolute, non-empirical relation of simultaneity. These responses need not be different, of course. Properly disambiguated, for instance, the different referents of `time' on the first view entails different referents of `simultaneity' and in this sense the first view entails the third view held by the earlier Prior. Conversely, however, one could certainly maintain the third view without any commitment regarding the first view. 

One who defends the third view without apparent commitment to the first view is John \citet{luc89}. Lucas also maintains that presentism does not violate any of the empirical consequences of SR and is thus compatible with it by pointing out that ``[t]he divine canon of simultaneity implicit in the instantaneous acquisition of knowledge by an omniscient being'' (220) is perfectly compatible with SR, as there may be ``a divinely preferred frame of reference'' (ibid.).\footnote{A later incarnation of Lucas, found in \citet{luc99}, defends an incompatibilist version of a similar idea by affirming an in principle observable preferred frame. I will return to this in \S\ref{sec:incom}.} Theology aside, the idea is to stipulate unobservable extra-structure in the form of an absolute simultaneity relation in order to satisfy an appetite dictated by a metaphysical agenda. Many presentists defend versions of this response, among them Ned \citet[\S3.9]{mar04} and Dean \citet{zim08}, even though Markosian's stance is less committal concerning what exactly SR does or does not entail. In fact, Markosian only asserts the disjunction that either this third compatibilist view is correct or else SR entails that there cannot be such an absolute simultaneity relation, in which case, however, SR must be rejected on incompatibilist grounds and based on ``good {\em a priori} evidence'' (75).  Zimmerman accepts that SR encodes the geometry of spacetime, but denies that this entails any ontological consequences. In particular, nothing in SR prohibits an absolute non-empirical simultaneity relation whose existence Zimmerman asserts.

If this stipulation of extra-structure is motivated purely by a presentist metaphysics, we better have very good reasons for believing presentism. The usual justifications for presentism trade on intuitions allegedly grounded in common sense which are said to powerfully demand that only presently existing things really exist. I, for one, only have weak intuitions regarding these matters; so weak that they are easily trumped by reasoned argument. But suppose another philosopher's intuitions are so strong as to warrant this step. Still donning our naturalist hat, it seems odd that many humans would have evolved intuitions that must depend on a structure which cannot be detected in principle. So either philosophers overestimate the extent to which humans have intuitions of the requisite kind or else these intuitions do not ontologically depend on an ultimately unobservable extra-structure such as absolute simultaneity. Most likely, of course, these intuitions---to the extent to which we have them---arose as an adaptation useful for beings operating at human scales, with the slow motions predominant in our empirical world.

Returning to the first view, according to which SR and presentism simply talk about different things, the main problem it confronts is a tenacious charge of obscurantism: if the time presentism speculates about is distinct from that which SR, and physics quite generally, theorizes about, what then is it? The time of physics is that which is tracked by any physical clock, from atomic clocks to biological and astronomical ones. The presentist's time, on the other hand, cannot possibly find any expression in the physical realm; for if it did, we could observe its regularities and compare them to other physical ones. Unless it would show a violation of Lorentz invariance, however, these regularities would have to accord to the Lorentz symmetry postulated by SR and would thus lead the presentist back to the challenge as given in \S\ref{sec:chall}. In case it {\em did} violate Lorentz invariance, we would have found an empirical confutation of SR. This, in turn, would signal not a compatibility of presentism with SR, but that new physics was required. Indeed, the presentist would find herself in the top right box of those who resolve the tension by showing that while presentism is incompatible with SR, it is perfectly consistent with more fundamental, `better' physics. I shall turn to resolutions of this type in \S\ref{sec:incom}. 

Time, therefore, must remain obscure on this view.\footnote{That Priorian presentism trades on the obscure is ironic, given that \citet[160]{pri67} accused eternalists (or, more precisely, detensers) of ``superstition'' because they ``pretend not only to resurrect the dead but even to summon forth the unborn'', in the words of \citet[23]{mas69}.} Furthermore, Gerald \citet{mas69} accused Prior's programme of tense logic and presentism to be ``grounded in bad physics and indefensible metaphysics'' (31f). Yet, despite this, and arguably because of its many ingenious innovations  that even Massey acknowledged, Prior's presentism continues to be influential. Jonathan Lowe, in his contribution to this collection \citep{low12}, seems to defend a similar line to Prior in his first comment of Section III, where he insists that we only get from the merely operational definition of time as found in physics to the conclusion that this characterization does really track {\em time} by additional metaphysical premises. These ancillary assumptions needed for the interpretation of the formal theoretical structure of SR can be chosen in different ways; in particular, Lowe maintains, they can be chosen as to permit metaphysical systems with absolute time and absolute simultaneity. Again, it is hard to see how this resolution avoids obscurantism.

What Lowe really believes is that a presentist can accept SR's stricture that there cannot be absolute simultaneity, as his further comments show. The way to evade the grip of the challenge, for Lowe, is to deny that co-existence is an equivalence relation. In the classification scheme proposed here, this resolution falls under the second compatibilist view, which now remains to be discussed. There are, of course, many ways of modifying presentism such as to keep it in line with SR. One important group of modifications denies that co-existence is an equivalence relation. As violating reflexivity is not attractive, approaches in this camp either deny symmetry or transitivity. Lowe proposes to violate transitivity, as he believes the demand that co-existence is transitive constitutes a metaphysical assumption motivated by an eternalist understanding of temporal reality that the presentist naturally rejects. As he most specifically explicates in comment (c) in Section III, transitivity ought to be rejected because on the endurantist conception of persistence he maintains, a person is wholly present at all times he or she exists, and co-exists with particular tropes of hers at each of these times, while the tropes of hers instantiated at different times do not co-exist.\footnote{A {\em trope} is a particular instance of a property or a relation, holding of, or co-existing with, the concrete particular it characterizes.} This last point is a distinctively presentist thesis, and insisting on the transitivity of co-existence amounts to an eternalist prejudice in that it is assumed that the tropes at different times co-exist---or so he claims.

Let me first note how unpalatable giving up the transitivity of co-existence really is. Without transitivity, it seems impossible to have a determinate and objective fact of the matter as to what the sum total of existence is. Existence seems relativized if I have to accept that what exists relative to $b$ may not exist relative to $a$, even if $b$ exists relative to $a$. We will soon see a more radical version of a proposal along these lines, but I find Lowe's proposal unattractive because it appears to have significant costs in the currency of the objectivity of existence, while it may not resolve the difficulties originating in SR at all. The reason why I am sceptical of its efficacy of eliminating the tension is that it seems as if at the level of {\em tropes}, transitivity is still required. If so, the problem returns in an unmitigated form. 

But why would the transitivity of co-existence be necessary for tropes? Suppose there are three people, Alice, Bob, and Carol, pairwise at some spatial distance from one another, but mutually at rest. In the metaphysical picture drawn by Lowe, all three constitute a series of tropes with which they co-exist, sequentially, at subsequent times. This much seems unproblematic, since for one (idealized) observer, time forms a total order even in a relativistic context. But now it seems as if there ought to be a fact of the matter which `Alice'-tropes co-exist with which `Bob'-tropes, etc. Suppose that the Alice-trope $A$ co-exists with the Bob-trope $B$ and that the Bob-trope $B$ co-exists with the Carol-trope $C$. Does $A$ co-exist with $C$? If so, without loss of generality, it seems as if at least within the same frame of reference (remember that the Alice, Bob, and Carol are mutually at rest), co-existence ought to be transitive if the relata are tropes---at least there seems no reason why it shouldn't be. Idealizing the tropes as being located at one spacetime point, SR mandates that if $A$, $B$, and $C$ are simultaneous (and hence co-exist) in the rest frame of Alice, Bob, and Carol, then they will not be simultaneous in any other frame. And it is not the case that this rest frame is privileged---it could have been the case that the three observers move relative to one another. Suppose that Bob and Carol start to move relative to Alice and relative to one another, even though they still move inertially. So within Alice's rest frame, which is the same rest frame as the one we had before and for which we established transitivity. But since Alice is in no way preferred over Bob or Carol, the same should be true for their rest frames. Hence, within each observer's rest frame, the transitivity of co-existence with tropes as relata should be valid. It may, however, not obtain {\em across different} frames, at least there is nothing in SR which would decide this matter. Either it does or it doesn't. If it does, then we are back to square one and the challenge still stands; if it does not, then the presentist accepts that existence gets fragmented and relativized to reference frames. The presentist of a Lowean persuasion finds himself between a rock and a hard place.

None of this should suggest that a presentist couldn't respond to the challenge by accepting the lessons of SR and, accordingly, relativize existence to inertial frames, as does Kit \citet[\S 10, 298-307]{fin05}. Simultaneity, and hence co-presentness, is defined only relative to an inertial frame. Therefore, and since for the presentist existence is tied to co-presentness, existence becomes fragmented in that it is only determinate with respect to a frame of reference. Co-existence is only an equivalence relation with respect to an inertial frame, but not {\em simpliciter}, as transitivity cannot act across frames. The price to be paid for this perfectly straightforward resolution, however, is immense: it requires a radically new understanding of physical existence. On a standard conception of physical existence, I take it, what exists is independent not only of the subject, but also of its kinematic state. On Fine's view, what co-exists with me depends on how I move. Thus, if we meet in the street, leisurely walking towards one another, what co-exists with you is entirely and completely different from what co-exists with me (with the exception of the {\em here-now}). Fine insists that this is a feature of his view, not a bug; but it is a feature which seriously modifies our conception of physical existence. Many presentists, I would think, are unwilling to follow Fine in this radical step. 

There are other ways to deny that co-existence, or co-presentness, is an equivalence relation. One is suggested (but not ultimately defended) by Howard \citet{ste91} and could be termed {\em past-lightcone presentism}. Past-lightcone presentism consider all and only events on the past lightcone of the {\em here-now} as co-existing with the {\em here-now}. Clearly, co-existence thus becomes non-symmetrical, as is evident in Figure \ref{fig:stein}: $q$ co-exists with $p$ but not vice versa.
\begin{figure}
\centering
\epsfig{figure=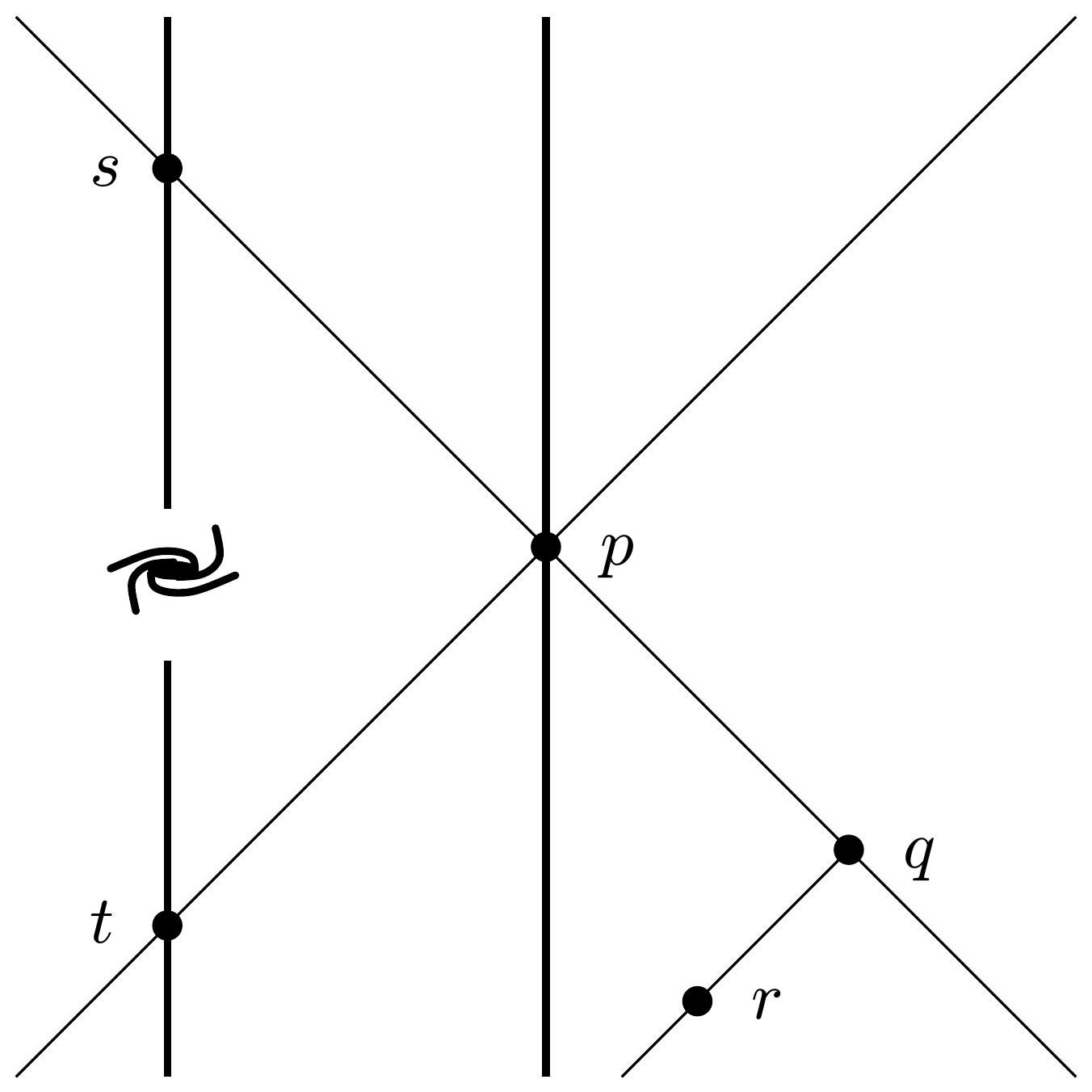,width=0.45\linewidth}
\caption{\label{fig:stein} An illustration of the violation of transitivity and of the insistence on symmetry in past-lightcone presentism}
\end{figure}
The loss of transitivity is also manifest: $r$ co-exists with $q$, and $q$ co-exists with $p$, but $r$ does not co-exist with $p$ because it is not {\em on} the past lightcone, but {\em inside} it. To save transitivity by including the full past lightcone of $p$ as co-existing with $p$ should not appeal to a presentist, unless she secretly harbours sympathies for the growing block view. The symmetry, but not the transitivity, of co-existence can be restored by extending the set of events which stand in the co-existence relation to $p$ to include those events on the {\em future} lightcone. Thus, if $q$ co-exists with $p$, so does $p$ with $q$. Such an extension, however, would have one rather counterintuitive consequence (cf.\ Figure \ref{fig:stein}): for a distant galaxy at rest relative to an observer on earth, events $s$ and $t$ far apart in time (e.g.\ some four million years for Andromeda), but not in space, would both co-exist with the present event $p$ for the earth-bound observer. This seems a rather odd outcome for a presentist; I suspect that most past-lightcone presentists would therefore refrain from saving symmetry. 

Let's tally the benefits and costs of past-lightcone presentism. First, this species of presentism is clearly compatible with SR as it defines the metaphysically salient structure purely in the Lorentz-invariant terms of the lightcone structure of spacetime. As a second advantage, co-existence tracks epistemic accessibility: all those events co-exist with the {\em here-now} which can be causally connected to (but not from) the {\em here-now}, e.g.\ by a light signal; consequently, all those events co-exist with the {\em here-now} which can be seen, at the {\em here-now}, as occurring now. While these virtues are not insignificant, they are outweighed by the approach's problems. First, it is questionable to what extent it deserves the moniker `presentism' as it includes as co-existent with the {\em here-now} events reaching arbitrarily far back into the past. There are events all the way back to, but not including, the big bang which co-exist with the {\em here-now}.\footnote{\label{fn:flrw} Thus including events located at all cosmological times from arbitrarily close to the big bang to today. In the standard cosmological models of Friedmann-Lema\^{\i}tre-Robertson-Walker spacetimes, these cosmological times are privileged against all other ways of foliating these spacetimes; cf.\ \S \ref{sec:incom}.} Moreover, past-lightcone presentism requires the unjustified awarding of a privilege of the spatially present over other spatial locations. The imposition of a prerogative of the {\em here} is implicit in the position because of the unique role played by the apex of the lightcone. The past lightcones of two distinct spacetime points are generally distinct.\footnote{Except if the spacetime lacks a property called `past-distinguishing'.} Selecting one past lightcone as that which contains events enjoying an ontological distinction over the others thus means to {\em spatiotemporally} privilege a location---and not just {\em temporally} as the presentist routinely does. Space and time are thus treated much more on a par than may be usual, or desirable, for a presentist. While presentists go at great lengths offering a justification for distinguishing the present, the past-lightcone version of presentism would only be appealing, counterfactually, if similar justifications could be offered for the prerogative of the {\em here-now} as for the {\em now}. 

The one remaining compatibilist presentism I wish to discuss also modifies the original position, but in rather different ways. James \citet{har08} has defended a `pointilliste' version of presentism according to which not only is the sum total of existence restricted to the temporally present, but it is also limited to the {\em spatially} present. This point presentism evidently relies only on the Lorentz-invariant structure of relativistic spacetimes: single points. Thus, the challenge mounted in \S\ref{sec:chall} does not even get to the starting blocks. This solipsistic version of presentism, however, is a very lonely view indeed: not even all of me exists! Furthermore, solipsist presentism fails to capture the spirit of presentism, as \citet[304]{fin05} reminds us, which maintains that there is a metaphysically deep distinction between space and time in that there exists an objective `now', even though there is no equally objective `here'. Finally, and relatedly, a justification for privileging this rather than that spacetime point is required, just as it was for the past-lightcone presentism. 

Summarizing our findings of this section, we can safely reject the claim that SR precludes a presentist metaphysic. Presentism is not physically impossible according to SR. Even assuming the strict truth of SR, there are many ways in which a presentist can evade the pressure originating from the relativity of simultaneity. All of these ways, however, incur certain costs; costs that are, in my view, too high to justify any potential gain they might offer.

\section{Incompatibilism with special relativity}\label{sec:incom}

There is almost universal agreement that SR is not a true theory. It assumes the complete absence of gravity, for instance. Because gravity shapes the structure of spacetime, the Minkowski spacetime we find in SR cannot adequately describe the spacetime structure of the world we live in. Furthermore, SR does not take any quantum effects into consideration. If a naturalistically inclined presentist presented an argument from some physical theory better than SR that would establish that the challenge produced in \S\ref{sec:chall} would no longer go through, she would offer respite for presentism from the besieging relativity of simultaneity. Arguments of this type count as incompatibilist because they accept the verdict from \S\ref{sec:chall}, but try to overturn it by rejecting SR. It is important to note that SR is not simply rejected on a priori or otherwise antinaturalist grounds, but instead because it is believed to be an ultimately false theory of the actual world, to be replaced by a better theory. This naturalist assumption dictates the rules for this section: any presentist opting for this route must produce at least an equally good (interpretation of a) theory on which the problem vanishes, where `good' is judged by the standards of physical science. This means that the metaphysician must get her hands dirty and analyze some actual physics. Such an analysis very quickly leads into a thicket of foundational questions in special and general relativity, quantum mechanics, quantum field theory, and quantum gravity. Naturally, I cannot possibly cover all the possible physics on which such an incompatibilist argument could turn in the remainder of this essay, but I will try to give you a sense of where to look and how such an argument might go.

The physics invoked, or reinterpreted, by the naturalist incompatibilist is either classical, i.e.\ non-quantum, or it relies on some quantum effects to get around SR's stricture of the relativity of simultaneity. Today, I shall focus on some `classical' strategies and only briefly comment on some quantum considerations towards the end of the essay. Among those, either a reinterpretation of SR, or the identification of extra-structure in general relativity (GR)---a more fundamental theory than SR---are most promising. One rather popular strategy of the first type, found e.g.\ in William \citet{cra01}, seeks a `neo-Lorentzian' reinterpretation of SR. Emulating Hendrik Lorentz's postulation of an immobile aether, it introduces a preferred frame of reference just as the aether would have done. Often, however, proponents of this strategy desist from offering a particular physical mechanism such as the aether which would physically explain the privilege awarded to one frame in particular. The important point, of course, is that it is in principle impossible to detect the preferred frame. Mathematically speaking, this fact gets encoded in the strict validity of  Lorentz symmetry, which still obtains. This strategy is the identical twin of the compatibilist strategy of insisting that SR is compatible with the postulation of extra-structure as a preferred frame. The only difference here is that the standard reading of SR is assumed to prohibit such extra-structures, and hence rejected and supplanted by a neo-Lorentzian version which includes the preferred frame. Ultimately, whether this strategy is considered compatibilist or incompatibilist thus boils down to the issue of whether SR permits grafting on the extra-structure of a preferred frame, as we have already seen e.g.\ in the strategy employed by \citet{mar04}. I take no stance on this essentially semantic question but will henceforth use the term `SR' to include a prohibition of any preferred frames and `neo-Lorentzian SR' to designate SR-cum-preferred frame. 

Against the twin strategy of adding an absolute, non-empirical simultaneity it can be complained, as I did above, that it violates the demands of Ockam's razor by postulating excess entities whose effects cannot even in principle be detected. Apart from the charge that it relies on unnecessary entities, neo-Lorentzian SR seems to make the Relativity Principle mentioned in \S \ref{sec:chall} only accidentally true. While the Relativity Principle is of course not metaphysically necessary, let me emphasize that neo-Lorentzian SR retracts what many consider SR's major accomplishment, viz.\ to show that not only is the Relativity Principle a deep principle of fundamental physics, but it can consistently be maintained alongside another successful empirical generalization: the Light Postulate. Furthermore, standard SR and the view of spacetime it promulgates lend themselves---unlike neo-Lorentzian SR---quite naturally to the development in understanding spacetime brought about by GR. In the realm of GR, which liberally admits many spacetime geometries and even topologies such that, in general, spacetimes can no longer be carved up into slices of space ordered by time. Thus, in those worlds at least where such a foliation is not possible at all, we do not even {\em get} to the problem of having to privilege one frame among infinitely many for no good empirical reasons---there simply are no such global frames anymore. In other words, Neo-Lorentzian SR seems to exhibit all the vices of ad-hockery and none of the virtues of ex ante, testable explanations with independent support. Neo-Lorentzian theories are driven either by a refusnik attitude towards the lessons of SR or by some more explicit metaphysical agendas; either way, they make for bad physics. As long as we are constrained to the non-dynamical Minkowski spacetime, there is no good reason to adopt a neo-Lorentzian preferred frame. But new possibilities open up once the narrow confines of special-relativistic physics dehisce. 

Staying within the classical incompatibilist camp though, a popular strategy utilizes the cosmological models of GR to reintroduce and justify a privileged time and thus an absolute simultaneity. Motivated by the idea that no location in space, including ours, is physically privileged (the so-called {\em Copernican} or {\em Cosmological Principle}), cosmologists assume that a necessary condition for the Copernican Principle to hold is that spacetime is spatially homogeneous. A theorem \citep{wal44} establishes that a sufficient condition for spatial homogeneity is the exact spherical symmetry around every point of the spacetime. The theorem also shows that if the condition of exact spherical symmetry about every point is satisfied, then the spacetime can be foliated into spacelike hypersurfaces of constant curvature. Spacetimes which exhibit exact spherical symmetry about every point are the Friedmann-Lema\^{\i}tre-Robertson-Walker (FLRW) spacetimes mentioned in footnote \ref{fn:flrw}. The foliation into spacelike hypersurfaces they admit is unique in that for only one such foliation it is the case for each hypersurface that all points in it exemplify the same spatial curvature. The foliation is thus physically privileged, and the parameter which orders the folia is called {\em cosmological time} $t$. Thus, the FLRW spacetimes---the cosmological standard models---admit an absolute time and an absolute notion of simultaneity: two events are {\em FLRW-absolutely simultaneous} just in case they are within the same spatial hypersurface of the privileged foliation or, equivalently, occur at the same cosmological time $t$. This notion of simultaneity is absolute since for any two events in an FLRW spacetime it is either the case that they are FLRW-absolutely simultaneous or not.\footnote{For a more systematic account of FLRW spacetimes, cf.\ e.g.\ \citet[Ch.~5]{wal84}.}

The move from SR to GR thus seems to reinvigorate the naturalist presentist's enterprise. As already James \citet{jea36} recognized, with apparent relief, the FLRW spacetimes make ``a real distinction between space and time'', such that we have ``every justification for reverting to our old intuitional belief that past, present, and future have real objective meanings, and are not mere hallucinations of individual minds---in brief that we are free to believe that time is real.'' (23; cited after \citet[116f]{loc05}) Many presentists have followed Jeans in imbuing cosmological time with ontological significance. But this move is not without its shortcomings. Michael \citet[105]{ber89} resists the inference from the fact that there is a uniquely most natural reference frame for FLRW spacetimes---the one at rest with respect to the local matter of the universe averaged over vast distances---to the conclusion that there is absolute space and time. I concur with Berry, but let's consider some more specific problems of the Jeansian proposal. 

An immediate problem already noted by Kurt \citet[560n]{goe49a} is that relying on cosmological time to define absolute time seems to yield only an approximate definition. It can only provide such an approximation because the assumptions undergirding the FLRW spacetimes are idealizations; of course, our actual universe is embarrassingly obviously {\em not} spatially homogeneous. In fact, it is hard to imagine how life would be possible in a perfectly homogeneous universe. So at small scales, we find blatant inhomogeneities. The question thus arises at which scales the idealizing assumption of spatial homogeneity is valid within the limits demanded of the approximation. This is G\"odel's point: at no scale smaller than the full universe have different spatial regions in general the exact same average spatial curvature; thus, to make the definition precise, either nothing short of the full universe will work, or else arbitrary elements ``such as the size of the regions or the weight function to be used in the computation of the mean motion of matter'' (ibid.) must be introduced. Judging from this, G\"odel found it ``doubtful whether there exists a precise definition which has so great merits, that there would be sufficient reason to consider exactly the time thus obtained as the true one.'' (ibid.)

To use an analogy from Michael \citet[118]{loc05}, just as the surface of earth is idealized as a perfect sphere, or as a perfect oblate spheroid, when in reality it is, at least from up close enough, a rocky asteroid, the hypersurfaces of FLRW spacetimes are idealized to be perfectly homogeneous when in actuality they are, at least from up close enough, rather inhomogeneous. The equivalence classes of FLRW-absolutely simultaneous events thus rely on an idealized division of space and time which may locally well be violated. In all this, it remains utterly mysterious how this highly idealized construction connects to our intuitions regarding temporal becoming and the present. If the absolute time constructed from this idealizing averaging procedure over vast cosmic scales is the time which determines what is the present, then how does the human perceptual and cognitive apparatus latch on to this idealized structure? In order for us to have truthful intuitions regarding the present, as a necessary condition, there must be a causal story of how humans pick up the present so defined. There are reasons to believe that such a causal mechanism cannot operate even in principle---after all, the spacelike hypersurfaces of constant spatial curvature which define the present extend across all of the universe and include parts from where light signals can only reach earth in a few billion years. Clearly, our presentist intuitions, should we have them, must be generated in a different way.

That this will not be trivial to resolve can be gleaned from explicating a useful distinction between `public' and `private' spaces made by Wolfgang \citet{rin81}.\footnote{The distinction originated in Edward A Milne's discussion of his eponymous spacetime, but was generalized by \citet{rin81}. I thank David Malament for teaching this material to me.} Both public and private spaces are spacelike hypersurfaces of a four-dimensional general-relativistic spacetime. Consider an infinite number of test particles whose trajectories are timelike geodesics. A {\em private} space then is  a spacelike hypersurface generated by (spacelike) geodesics which are orthogonal to the timelike curve of a particular test particle. This is, at it were, the test particle's own private `space', viz.\ the space orthogonal to its `time'. A {\em public} space, on the other hand, is a spacelike hypersurface which is everywhere orthogonal to a family of timelike curves. Restricting ourselves to the case of `open' FLRW spacetime (i.e., the spatial curvature at events in the hypersurfaces of constant spatial curvature is non-positive), the so-called Penrose diagram given in Figure \ref{fig:penrose} gives a graphical illustration of the difference between private and public space. 
\begin{figure}[htb]
\centering
\resizebox{0.75\textwidth}{!}{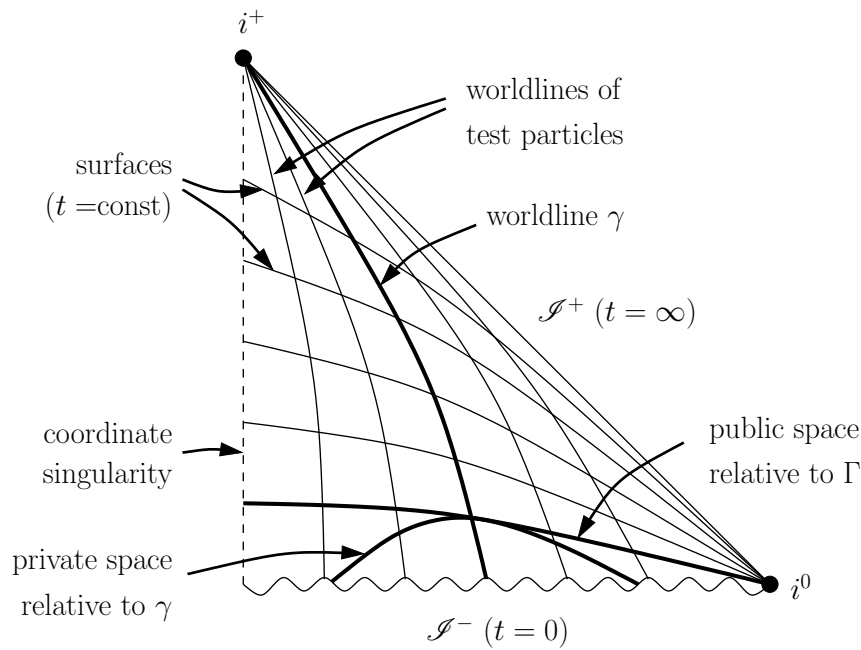}
\caption{\label{fig:penrose} Penrose diagram of an open FLRW spacetime (with $p=\Lambda=0$) with a family $\Gamma$ of worldlines of test particles}
\end{figure}
A Penrose diagram represents the conformal structure of a spacetime, i.e., a way of representing the structure of an infinitely extended spacetime in a finite diagram. The straight boundary lines represent infinity, the wavy boundary lines a singularity, the dashed boundary lines symmetry axes or coordinate singularities, and points points. Boundary null surfaces are labelled $\mathscr{I}$ (read `scri'), with $\mathscr{I}^+$ and $\mathscr{I}^-$ representing future and past null infinity, respectively, and boundary points $i$, with $i^+$ and $i^-$ designating future and past timelike infinity, respectively, and $i^0$ spacelike infinity. In Figure \ref{fig:penrose}, any timelike geodesic originates in $\mathscr{I}^-$ and ends in $i^+$, and null geodesics start in $\mathscr{I}^-$ and finish in $\mathscr{I}^+$, and spacelike geodesics originate and end in $i^0$. (But non-geodesic curves do not follow these rules).

Let us consider a family $\Gamma$ of timelike geodesics representing the worldlines of test particles, as well as one particular representative $\gamma$ of that family (cf.\ Figure \ref{fig:penrose}). As indicated in Figure \ref{fig:penrose}, the public space relative to $\Gamma$ at a given time $t=t'$ is just the spacelike hypersurface $t=t'$ of constant spatial curvature of the cosmologically privileged foliation. The private space relative to $\gamma$, however, only intersects this public space at the very point which jointly belongs to the worldline $\gamma$ and the public space relative to $\Gamma$. It should be noted that both these spaces are supposed to represent `space' as it is given at the same time $t=t'$, just once with respect to a family of observers or test particles and once for just one observer. The two notions of space are clearly inequivalent. In particular, the private space curves back onto the initial singularity $\mathscr{I}^-$ of the `big bang', including arbitrarily early moments of cosmological time. A proposition by Don \citet{pag83} establishes that private space is finite in any homogeneous and isotropic general-relativistic cosmological spacetime which is expanding (and satisfies certain other conditions). 

Thus, the presentist is forced to disambiguate between the two notions of space. At most one of them correctly captures the structure of the spatially extended present. But which one to pick? Given that the presentist's original inclination was to utilize the cosmologically privileged foliation to re-introduce absolute simultaneity, the notion of public space seems more promising. It certainly commands the more objective validity in that it does not randomly, or at least unjustifiedly, select one worldline to fill a special role. In FLRW spacetimes, the public space also doesn't get arbitrarily close to the big bang, but instead tracks a more natural notion of simultaneity. Public space is only well-defined in spacetimes which do not rotate or, equivalently, for which there are families of worldlines such that the spacetime can be foliated into a family of spacelike hypersurfaces which are orthogonal to the worldlines.\footnote{Cf.\ David Malament, `How space can be (and is) finite', talk at UCSD on 8 June 2009.} Thus, the infamous G\"odel spacetime does not permit public spaces. Perhaps this is not a big loss for the public-space presentist; but it does make presentism vulnerable to non-standard spacetime structures, which may well be actual for all we know. 

Would it be safer for the presentist to bet on private space as encoding the structure of the spatially extended present? After all, this seems what a finite, earthbound observer could hope to construct. It can often be constructed even in spacetimes in which no well-defined public space exists, such as in the rotating universes which fail to be hypersurface orthogonal. But we need not go far to recognize the weaknesses such a private-space-based approach would have. It is evidently egocentric as distant observers will never agree on what the present is, just as in the case of solipsist and past-lightcone presentisms. Strictly speaking, you and I will always disagree about which events are present. Symmetry (and transitivity) is lost again, as those temporal parts of you which are real according to {\em me-now} take a temporal part of me to be real which is in the causal past of the {\em me-now}. Since different observers have different private spaces, and only one of them gets the ontological privilege, a justification for singling out this, but not that, observer is again required. It is hard to see {\em how} such a justification could be forthcoming. Ironically then, private-space presentism unduly awards an ontologically special status to earthbound observers after having relied on modern cosmology whose Cosmological Principle exactly {\em denies} any special status to {\em us-now}. 

There are two further repulsive features of private-space presentism. First, private spaces are not in general extendible to universal spaces, i.e., even for some causally benign spacetimes, they do not intersect the worldlines of all observers, so that some observers have no temporal parts which are ever real.\footnote{More accurately, there exist private spaces in globally hyperbolic spacetimes which are not Cauchy surfaces.} Second, and just as for past-lightcone presentism, in FLRW spacetimes part of an observer's private space will always be arbitrarily close to the big bang. This seems hard to reconcile with the presentist intuition that it is the present, not the past, that deserves the noble epithet of reality. 

In sum, FLRW spacetimes offer a much less hospitable venue to a presentist metaphysic as the incompatibilist presentist may have hoped. But not all classical hope is lost. As Bradley \citet{mon06} has reminded us, GR contains a large class of spacetimes which seem amenable to a principled procedure for introducing unique foliations into space and time, one that even avoids the gross idealizations that paved the way to cosmological time. This procedure slices the four-dimensional spacetime into spacelike hypersurfaces parametrized by constant mean (extrinsic) curvature, or CMC.\footnote{This curvature is defined as the trace of the extrinsic curvature, i.e., of a mathematical magnitude which quantifies how the three-dimensional hypersurface is embedded into the four-dimensional spacetime. It thus differs from the purely three-dimensional, and hence intrinsic,`spatial' curvature utilized in the introduction of the cosmological time in FLRW spacetimes.} I will spare you a detailed assessment of the prospects of presentism based on CMC foliations, as I have given one in \citet{wut10}. But my conclusions there were negative: apart from numerous technical problems and from the callowness of the approach, the most devastating problem was even if the large-scale structure of our actual universe is best described by a spacetime which admits a CMC foliation, and even if one of the folia of this CMC foliation is rightly distinguished as the present, it remains far from clear, to put it mildly, how it can be that it is this CMC foliation that our presentist intuitions are tracking. Why should our sense that the present is somehow ontologically special be sensitive to the constant mean extrinsic curvature of spacelike hypersurfaces? Clearly, it is not enough to simply identify a folium of a certain constant mean curvature as the present and believe that one has explained our presentist intuitions.

While this seemingly exhausts at least the most obvious and the most viable classical strategies available to the incompatibilist presentist, many presentists have turned to quantum physics and have drawn new hope from several aspects of the quantum. Doing them the justice they deserve will have to wait for another day, so let me just list the two most obvious routes that have been pursued, with a few quick comments. They both concern particular interpretations of non-relativistic quantum mechanics, Bohmian mechanics and collapse theories. As an example of utilizing the latter, John \citet[10]{luc99}---a later temporal part of the compatibilist mentioned in \S\ref{sec:com}---offers a forceful statement of how collapse interpretations provide a home for a physically distinguished present, adorned with the temporal asymmetry so beloved by presentists:
\begin{quote}
There is a worldwide tide of actualization---collapse into eigenstate---constituting a preferred foliation by hyperplanes (not necessarily flat) of co-presentness sweeping through the universe---a tide which determines an absolute present [...] Quantum mechanics [...] not only insists on the arrow being kept in time, but distinguishes a present as the boundary between an alterable future and an unalterable past.
\end{quote}
If the collapses invoked by Lucas are to be real physical mechanisms---which they would have to be in order to fill the role assigned to them by collapse presentists---, then they occur in a particular basis. For instance, in a GRW collapse theory, the collapses occur in the position basis. Whichever basis the collapse presentist chooses, her selection must be given a physical justification. I have no reason to assume that this can't be done, but would like to emphasize that it does not suffice to simply invoke collapse as a physical mechanism to distinguish the present and leave it at that. Furthermore, the collapses' blatant violation of Lorentz symmetry is usually regarded by physicists not as a metaphysical virtue, but as a physical vice. Therefore, physicists are searching for a relativistic version of collapse interpretations such as GRW. Such relativistic collapse theories should be expected to no longer rely on a preferred foliation of spacetime, but instead to collapse the wave function in a Lorentz-invariant way.\footnote{This is indeed what happens in the only current candidate for such a theory, \citet{tum06}'s `rGRWf'.} In fact, given alternative proposals to solve the measurement problem in quantum mechanics such as Everettian many-worlds theories and hidden-variables theories such as Bohmian mechanics, it is evident that quantum mechanics does not require collapse at all. Among those working in the foundations of quantum mechanics today, I would estimate that only a minority advocates collapse interpretations. The rejection of collapse interpretations, of course, does not entail an impossibility for the presentist to find a physical structure incarnating her metaphysical fantasy. Perhaps Bohmian mechanics, or non-local Bell correlations, or the quantization of spacetime, offers an attractive route to its fulfillment. But this, as I said, is the topic for another occasion.

As a general reminder to compatibilist and incompatibilist presentists alike, let me finish by stressing that the strictures of SR are quite strong; Lorentz symmetry is fantastically well confirmed in many disparate contexts and for many different phenomena.\footnote{For an authoritative recent review of the main standard tests of Lorentz symmetry, cf.\ \citet{wil05a,wil05b}. \Citet{saleal08} have tested for a privileged frame in the context of non-local Bell correlations and found no indication that there is any.} As a consequence of this high degree of experimental and observational confirmation, it would be rational to expect Lorentz symmetry to be part of the true fundamental theory---although there is admittedly more to be said here about possible high-energy corrections of exact Lorentz symmetry. Rather than as a theory which has been supplanted by GR, relativistic quantum field theory, and---ultimately---a quantum theory of gravity, we should regard SR as a `second-order constraint' on these more fundamental theories, as I have explicated in \citet[\S4]{wut10}. Quite generally, presentists often underestimate the dialectical work that needs to be done to get around SR's ruling that simultaneity is relative.

\section{Taking stock: the grim prospects of presentism}\label{sec:stock}

In conclusion, we have found that fundamental physics does not uniquely determine the metaphysics of time, and hence does not entail the denial of presentism. But it does impose constraints which any naturalist worth her salt must respect. Metaphysics need not be subservient to physics, but to completely ignore pertinent experimental findings and theoretical insights coming from the sciences testifies to philosophical {\em hybris} {\em par excellence}. It is worthwhile to recall that the naturalism that I have asked the reader to adopt is rather mild: it simply demands that no physically possible worlds are metaphysically impossible, where physical possibility gets judged by our best physical theories. Once we engage in a detailed analysis of just what it is that our best physical theories state as possible, we recognize that maintaining presentism, while defensible along many routes, bears a high cost. Most of this essay has been concerned with detailing that bill. 

While the costs are high along all routes, the toll they extract may be quite different---presentists get to pick among many different poisons. But both sides of the balance sheet must be considered, costs as well as gains. To give a detailed analysis of the real or alleged gains in adopting presentism remains beyond the scope of this essay, but they surely include a claimed accordance with our intuitions, in particular in that it seems to make sense of the apparently so prevalent becoming and and ever-present transience in our world. It is this dynamical `umph', this whoosh, that presentists often cite as their main explanatory accomplishment. 

The lesson I wish to draw from my analysis is that the tension between modern physics and presentism can be resolved, but that all resolutions either require unpalatable metaphysics or speculative science, which our best current knowledge cannot support. On the first option, the presentist position may become so disfigured as to more than offset any advantage that may have been gained by its accordance with our intuitions. Finally, it should be noted that in order for this claimed advantage of presentism to come into play at all, the presentist must identify the {\em physical} structure which could justifiably play the ontologically special role and the mechanism which explains how our temporal intuitions arise from this physical structure. After all, the presentist draws an inference to the best explanation from our intuitions to the fundamentally privileged ontological status of the present. Hence, however this present is characterized, there better be an account of what it is and how it causally affects us in a way as to give rise to our temporal experiences. And there seems to be no hope of delivering such an account if either the structure identified as the present or the causal mechanism are not physically tractable. To explicate this story is a tall order for both compatibilist and incompatibilist presentists. 

I submit, therefore, that modern physics renders the prospects of presentism quite grim. As this essay has shown, however, presentism ought to be of interest not just for the metaphysician, but also for the philosopher of physics, as its analysis cuts deep into the foundational meat of many a physical theory. 

\bibliographystyle{plainnat}
\bibliography{/Users/christian/Professional/Bibliographies/presentism}

\end{document}